\documentclass[11pt]{article}
\usepackage{graphicx}        

\def\lbabar{\mbox{{\large\sl B}\hspace{-0.4em} {\normalsize\sl A}\hspace{-0.03em}{\large\sl B}\hspace{-0.4em} {\normalsize\sl A\hspace{-0.02em}R}}}
\usepackage{relsize}
\def\babar{\mbox{\slshape B\kern-0.1em{\smaller A}\kern-0.1em
    B\kern-0.1em{\smaller A\kern-0.2em R}}}

\def\Kbar  {\kern 0.2em\overline{\kern -0.2em K}{}}

\def\Kp    {\ensuremath{K^+}}

\def\Kstarz  {\ensuremath{K^{*0}}}

\def\Kzb   {\ensuremath{\Kbar^0}}
\def\KzKzb {\ensuremath{K^0 \kern -0.16em \Kzb}}

\def\Dbar  {\kern 0.2em\overline{\kern -0.2em D}{}}

\def\Dzb   {\ensuremath{\Dbar^0}}
\def\DzDzb {\ensuremath{D^0 {\kern -0.16em \Dzb}}}

\def\Bz    {\ensuremath{B^0}}

\def\Bbar  {\kern 0.18em\overline{\kern -0.18em B}{}}

\def\Bzb   {\ensuremath{\Bbar^0}}
\def\Bu    {\ensuremath{B^+}}

\def\BzBzb {\ensuremath{B^0 {\kern -0.16em \Bzb}}}

\mathchardef\Upsilon="7107
\def\Y#1S{\ensuremath{\Upsilon{(#1S)}}}% no space before {...}!

\mathchardef\Deltares="7101
\mathchardef\Xi="7104
\mathchardef\Lambda="7103
\mathchardef\Sigma="7106
\mathchardef\Omega="710A
\def\Deltabar   {\kern 0.25em\overline{\kern -0.25em \Deltares}{}}
\def\Lbar {\kern 0.2em\overline{\kern -0.2em\Lambda\kern 0.05em}\kern-0.05em{}}
\def\Sigbar{\kern 0.2em\overline{\kern -0.2em \Sigma}{}}
\def\Xibar{\kern 0.2em\overline{\kern -0.2em \Xi}{}}
\def\Obar{\kern 0.2em\overline{\kern -0.2em \Omega}{}}
\def\Nbar{\kern 0.2em\overline{\kern -0.2em N}{}}
\def\Xbar{\kern 0.2em\overline{\kern -0.2em X}{}}

\def\BR{{\ensuremath{\cal B}}}

\def\ev   {\ensuremath{\rm \,e\kern -0.08em V}}
\def\kev  {\ensuremath{\rm \,ke\kern -0.08em V}} 
\def\mev  {\ensuremath{\rm \,Me\kern -0.08em V}} 
\def\gev  {\ensuremath{\rm \,Ge\kern -0.08em V}} 
\def\gevc {\ensuremath{{\rm \,Ge\kern -0.08em V\!/}c}} 
\def\tev  {\ensuremath{\rm \,Te\kern -0.08em V}}
\def\mevc {\ensuremath{{\rm \,Me\kern -0.08em V\!/}c}} 
\def\gevcc{\ensuremath{{\rm \,Ge\kern -0.08em V\!/}c^2}} 
\def\mevcc{\ensuremath{{\rm \,Me\kern -0.08em V\!/}c^2}}

\def\mus  {\ensuremath{\rm \,\mus}}

\def\mus        {\ensuremath{\,\mu{\rm s}}}    %% microsecond
\def\gsim{{~\raise.15em\hbox{$>$}\kern-.85em
          \lower.35em\hbox{$\sim$}~}}
\def\lsim{{~\raise.15em\hbox{$<$}\kern-.85em
          \lower.35em\hbox{$\sim$}~}}

\def\to                 {\ensuremath{\rightarrow}}

\def\pep2{PEP-II}

\providecommand{\eqref}[1]{Eq.~(\ref{eq:#1})}

\def\jetset74   {\mbox{\tt Jetset \hspace{-0.5em}7.\hspace{-0.2em}4}}

\long\def\inst#1{\par\nobreak\kern 4pt\nobreak
    {\it #1}\par\vskip 10pt plus 3pt minus 3pt}
 
\begin{document}
{\pagestyle{empty}
 
\begin{flushright}
SLAC-PUB-8684\\
\babar-PROC-00/25 \\
%\babar-PUB-\BABARPubYear/\BABARPubNumber \\
%hep-ex/\LANLNumber \\
October, 2000 \\
\end{flushright}
 
%% originally was 4 cm
\par\vskip 2.5 cm

% Title of the paper
\begin{center}
\Large {\bf Search for {\boldmath $B^+ \rightarrow K^+ \ \ell^+ \ell^-$} and {\boldmath $B^0 \rightarrow K^{*0} \ \ell^+ \ell^-$}}
\end{center}
\bigskip
 
\begin{center}
\large
Natalia Kuznetsova\\
Universiry of California, Santa Barbara \\
Physics Department, University of California, Santa Barbara, CA 93106-9530, USA\\
(for the \lbabar\ Collaboration)
\end{center}
\bigskip \bigskip
 
% Abstract
\begin{center}
\large \bf Abstract
\end{center}
Using a sample of $3.7 \times 10^{6}$ $\Upsilon(4S) \rightarrow B\bar{B}$ events
collected with the \babar\ detector at the \pep2\ storage ring, we search for the electroweak penguin decays $B^+ \rightarrow K^+ e^+ e^-$, $B^+ \rightarrow K^+ \mu^+ \mu^-$, $B^0 \rightarrow K^{*0} \ e^+ e^-$, and $B^0 \rightarrow K^{*0} \ \mu^+ \mu^-$.  We observe no significant signals for these modes and set preliminary 90\% C.L. upper limits of
\begin{eqnarray*}
{\cal B}(B^+ \rightarrow K^+ e^+ e^-)  & <  & 12.5 \times 10^{-6}, \\
{\cal B}(B^+ \rightarrow K^+ \mu^+ \mu^-) &  < & \phantom{2}8.3 \times 10^{-6}, \\
{\cal B}(B^0 \rightarrow K^{*0} e^+ e^-) & < &  24.1  \times  10^{-6}, \\
{\cal B}(B^0 \rightarrow K^{*0} \mu^+ \mu^-) & < & 24.5 \times 10^{-6}. \\
\end{eqnarray*}                                                              

\vfill
\begin{center}
Contribued to the Proceedings of the DPF 2000
Meeting\\ of the Division of Particles and Fields of the American Physical Society,\\
8/9/2000---8/12/2000, Columbus, Ohio
\end{center}
 
\vspace{1.0cm}
\begin{center}
{\em Stanford Linear Accelerator Center, Stanford University,
Stanford, CA 94309} \\ \vspace{0.1cm}\hrule\vspace{0.1cm}
Work supported in part by Department of Energy contract DE-AC03-76SF00515.
\end{center}

\newpage
\pagestyle{plain}
\section{Introduction}    	%) A SECTION HEADING
\vspace*{-0.5pt}
\noindent
The rare decays $B \to K\ell^+\ell^-$ and $B\to K^*\ell^+\ell^-$, where $\ell$ is either an electron or muon, are highly suppressed in the Standard Model and are expected to occur via electroweak penguin processes.  Standard Model predictions indicate~\cite{bib:ali1} that ${\cal B}(B\to K\ell^+\ell^-)\approx 6\times 10^{-7}$, while ${\cal B}(B\to K^*\ell^+\ell^-)\approx 2\times 10^{-6}$. These processes provide a possible window into physics beyond the Standard Model, since new, heavy particles such as those predicted by SUSY models can enter the loops in the effective flavor-changing neutral current transitions~\cite{bib:buchalla} $b\to s$.

Experimentally, the small expected rates make searches for these modes difficult.  Searches from CDF~\cite{bib:CDF} and CLEO~\cite{bib:CLEO1} have so far yielded only upper limits.

In this paper, we report the results of a preliminary analysis to investigate the backgrounds and the ability of the \babar\ detector~\cite{bib:babar} to reject them.  We have analyzed an on-resonance data sample of 3.2 fb$^{-1}$, representing about a third of the current \babar\ $\Upsilon(4S)$ integrated luminosity.  The main goal of our study is to test the performance of a blind analysis in which the event selection is optimized without use of the signal or sideband regions in the data.  We analyze four charged particle decay modes: $B^+\to K^+ e^+e^-$, $B^+\to K^+ \mu^+\mu^-$, $B^0\to K^{*0} e^+e^-$, and $B^0\to K^{*0} \mu^+\mu^-$. In each case, we include the charge conjugate mode as well.
%\pagebreak

%\textheight=7.8truein
\setcounter{footnote}{0}
\renewcommand{\thefootnote}{\alph{footnote}}
%\vspace*{1pt}\textlineskip	%) USE THIS MEASUREMENT WHEN THERE IS
\section{Analysis Methods and Event Selection}
\noindent
We select event with at least 5 good quality tracks, of which two are leptons with lab frame momenta $p_e > 0.5\ {\rm GeV}/c$  (electrons) or $p_{\mu} > 1.0\ {\rm GeV}/c$ (muons).  Electrons and positrons are also required to pass the $\gamma$ conversions veto.  The  $B\to J/\psi \ K^{(*)}$ and $B\to \psi(2S) \ K^{(*)}$ events
have the same topology as our signal processes and must be removed
with great care, especially since bremsstrahlung can lower
the dielectron mass with respect to the $J/\psi$ or $\psi(2S)$ mass. We remove
events with dilepton masses consistent with the $J/\psi$ or $\psi(2S)$, and we
apply a correlated cut in the $\Delta E$ vs. $M_{\ell^+ \ell^-}$ plane to account
for possible bremsstrahlung or track mismeasurement.
%  The $B\to J/\psi \ K^{(*)}$ and $B\to \psi(2S) \ K^{(*)}$ background (which has the same topology as the signal) is suppressed by removing events with dilepton masses consistent with those of the $J/\psi$ or $\psi(2S)$, as well as applying a correlated selection in the  $\Delta E$ vs. $M_{\ell^+ \ell^-}$ plane to account for effects of bremsstrahlung and track mismeasurement.  
The $B \rightarrow J/\psi \ K$ modes can also pass this veto if the kaon is misidentified as a lepton (most often a muon).  In a similar way $B^- \rightarrow D^0 \pi^-$, where $D^0 \rightarrow K^- \pi^+$, can pass our selection criteria if both of the leptons are fake.   These effects can be suppressed by re-assigning the particle masses and excluding mass combinations around the ${J/\psi}$ and the ${D^0}$.  Continuum background is suppressed by using a four-variable Fisher discriminant.  Finally, the signal region is defined as a rectangle in the plane defined by the beam-energy substituted mass of the $B$ candidate $m_{ES}$ and the energy difference~\cite{bib:babar} $\Delta E$: 5.272 $<$ $m_{ES}$ $<$ 5.286 GeV/$c^2$ (3$\sigma$) and $-0.10 < \Delta E < 0.06$ GeV ($|\Delta E| < 0.06$ GeV) for the electron (muon) channels.
%, where
%\vskip -1.4in   
%\begin{eqnarray*}
%m_{ES}&=&\sqrt{\left(\frac{\sqrt{s}}{2}\right)^2 - \left(\sum_{\alpha=1}^{n} {\bf p}^\ast_{\alpha}\right)^2 }, \\
%\Delta E&=& \sum_{\alpha=1}^n \sqrt{ m_{\alpha}^2 + |{\bf p}^\ast_{\alpha}|^2 }-\sqrt{s}/2.
%\end{eqnarray*}
%\vskip -1.2in   
%The index $\alpha$ is over the particles that make up the candidate $B$ meson system.  $m_{\alpha}$ are the masses of the particles and ${\bf p}^\ast_{\alpha}$ are their momenta measured in the $\Upsilon(4S)$ center of mass frame. $\sqrt{s}$/2 is one half of the center of mass energy. The signal box is defined in the following way: 5.272 $<$ $m_{ES}$ $<$ 5.286 GeV (3$\sigma$) and $-0.10 < \Delta E < 0.06$ GeV ($|\Delta E| < 0.06$ GeV) for the electron (muon) channels.                    
For the  $B^0\rightarrow K^{\ast0} \ell^+\ell^-$ channels, we reconstruct the $K^{\ast0}$
 in the $K^+ \pi^-$ final state.  The kaon candidate is required to be identified as a kaon, 
while there are no particle identification requirements on the pion candidate.  The mass of the 
$K^+\pi^-$ pair is required to be within 75 MeV/$c^2$ of the $K^{\ast0}$ mass.

\section{Physics results}
\noindent
Figure~\ref{fig:de_vs_Bmass_data} shows a large $\Delta E$ vs.~$m_{ES}$ region (the ``grand sideband'') and a small box
indicating the signal region for each of the four modes.     
Table~\ref{tab:results} lists the signal efficiencies, total yield,
the expected background, and the 90\% C.L. upper limits on the branching fractions.  The signal efficiencies were determined from the signal Monte Carlo events.  The efficiencies include the branching fractions for the $K^{\ast 0}$ modes.  Note that even though we carry through a background estimation procedure using the sidebands in data, we are setting an upper limit assuming that each event in the signal region is potentially due to the signal process.  The table also lists the total systematic error. The dominant systematic uncertainty is on the tracking efficiency (2.5\% per track).  

%%%%%%%%It is important to note that because we decided before the measurement that our result would be expressed as an upper limit, we do not need to apply the Feldman-Cousins procedure (Ref.~\cite{bib:feld}), for a two-sided confidence interval, but instead use the original procedure specified in the Particle Data Book~\cite{bib:PDG}, in which zero observed events corresponds to an upper limit of 2.3 events. 
 
\begin{figure}[!tb]
\begin{center}
\includegraphics[height=5.7in]{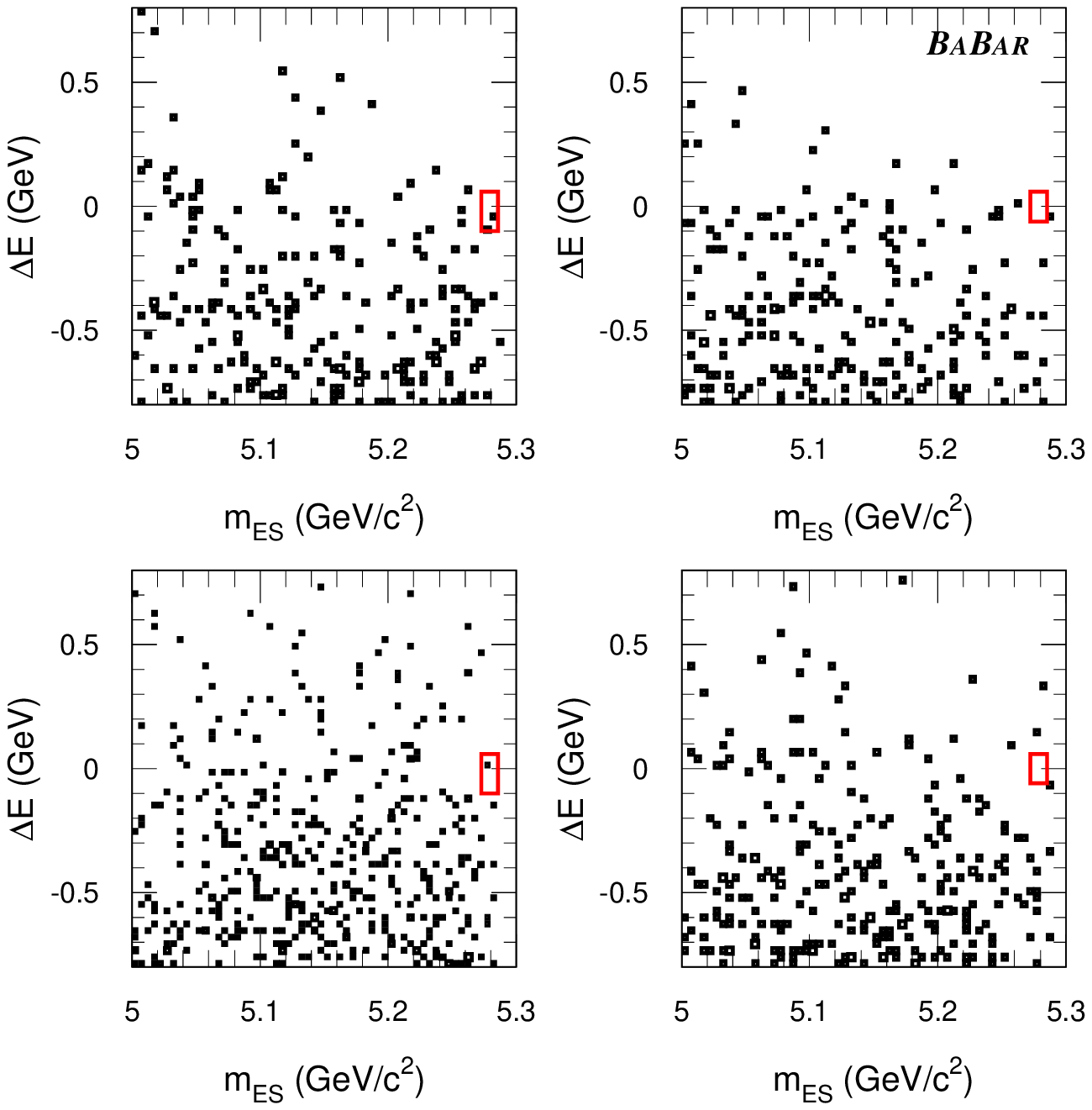}
\caption{$\Delta E$ vs. $m_{ES}$ (grand sideband) for data: (a) $\Bu\to \Kp e^+e^-$, (b) $\Bu\to \Kp \mu^+\mu^-$, (c) $\Bz\to \Kstarz e^+e^-$, and (d) $\Bz\to \Kstarz \mu^+\mu^-$.  The smaller boxes show the signal region.}
\label{fig:de_vs_Bmass_data}
\end{center}
\end{figure}

\begin{table}[!htb]
\caption{Signal efficiencies, systematic uncertainties (combining the
uncertainties on the signal efficiencies and on the number of produced
$\Upsilon$(4S) mesons), the number of
observed events, the number of estimated background events, and upper limits
on the branching fractions. In computing the upper limits we have assumed
$\BR(K^{*0}\to K^+\pi^-)=2/3$.
}
\vskip 0.5 cm
\begin{center}
\begin{tabular}{|l|c|c|c|c|c|}
\hline
Mode & Efficiency (\%) & Total systematic & Observed & Bkgd. estimated & $\BR/10^{-6}$ \\
     &                 & uncertainty (\%) & events   & from data          &  (90\% C.L.)\\
\hline\hline
$B^+ \rightarrow
       K^+ e^+ e^-$             &  13.1 & 11.7 & 2 & 0.20  & $<$ 12.5  \\
$B^+ \rightarrow
       K^+ \mu^+ \mu^-$         &  8.6  & 12.3 & 0 & 0.25  & $<$ 8.3  \\
%$B^0 \rightarrow
%       \Kstarz e^+ e^-$      &  ?  & 14.2 & 1 & 0.50  & $<$ 24.1  \\
%$B^0 \rightarrow
%       \Kstarz \mu^+ \mu^-$  &  3.1  & 14.8  & 0 & 0.33 & $<$ 25.2  \\
$B^0 \rightarrow
       \Kstarz e^+ e^-$       &  7.7  & 14.2 & 1 & 0.50  & $<$ 24.1  \\
$B^0 \rightarrow
       \Kstarz \mu^+ \mu^-$   &  4.5  & 14.8  & 0 & 0.33 & $<$ 24.5  \\
 
\hline
\end{tabular}     
\end{center}
\label{tab:results}
\end{table}

\section{Conclusion}
\noindent
We have searched for rare $B$ decays $B \rightarrow K^{(\ast)} \ \ell^+ \ell^-$
in a sample of $3.7 \times 10^{6}$ $B\bar{B}$ events.  We find no
observable signal for any of the four modes considered, and set 
preliminary 90\% C.L. upper limits on the branching fractions:
\begin{eqnarray*}    
{\cal B}(B^+ \rightarrow K^+ e^+ e^-)  & <  & 12.5 \times 10^{-6}, \\
{\cal B}(B^+ \rightarrow K^+ \mu^+ \mu^-) &  < & \phantom{2}8.3 \times 10^{-6}, \\
{\cal B}(B^0 \rightarrow K^{\ast0} e^+ e^-) & < &  24.1  \times  10^{-6}, \\
{\cal B}(B^0 \rightarrow K^{\ast0} \mu^+ \mu^-) & < & 24.5 \times 10^{-6}. \\
\end{eqnarray*} 
The limits for the $B^+\rightarrow K^+ \ell^+ \ell^-$ modes are comparable to those set by other experiments, 
while those for $B^0 \rightarrow K^{\ast0}\ell^+\ell^-$ are less sensitive with this 
data sample. We plan to analyze substantially more data in the near future.

%\pagebreak

\end{document}